\documentclass[twocolumn,secnumarabic,amssymb, nobibnotes, aps, prd]{revtex4-1}
\usepackage{graphicx}% Include figure files
\usepackage{dcolumn}% Align table columns on decimal point
\usepackage{bm}% bold math
\usepackage{caption}
\usepackage{subcaption}
\usepackage[]{algorithm2e}
\usepackage{stmaryrd}
\usepackage{graphicx}
\usepackage[latin1]{inputenc}
\usepackage{times}
\usepackage{tikz}
\usepackage{amssymb,amsbsy,amsmath,amsfonts,amssymb,amscd}
\usepackage{float}
\usepackage{verbatim}
\graphicspath{{figures/}}

\begin{document}

%\preprint{APS/123-QED}

\title{Lyapunov spectrum of separated flows and its dependence on numerical discretization}% Force line breaks with \\
%\thanks{A footnote to the article title}

\author{P. Fernandez}
% \altaffiliation[Also at ]{Department of Aeronautics and Astronautics, Massachusetts Institute of Technology, 77 Massachusetts Avenue, Cambridge, MA 02139, USA.}%Lines break automatically or can be forced with \\
\email{pablof@mit.edu}
\author{Q. Wang}%
 \email{qiqi@mit.edu}
\affiliation{%
 Department of Aeronautics and Astronautics, Massachusetts Institute of Technology, 77 Massachusetts Avenue, Cambridge, MA 02139, USA.
}%
%\author{P. Blonigan}%
% \email{pjblonigan@gmail.com}
% \affiliation{%
% NASA Ames Research Center, Moffett Field, CA 94035, USA.
%}%


%\affiliation{
%\affiliation{
%\affiliation{%

\date{\today}% It is always \today, today,
             
%\maketitle
%\tableofcontents

\begin{abstract}
We investigate the Lyapunov spectrum of separated flows and their dependence on the numerical discretization. The chaotic flow around the NACA 0012 airfoil at low Reynolds number and large angle of attack is considered to that end, and $t$-, $h$- and $p$-refinement studies are performed to examine each effect separately. Numerical results show that the time discretization has a small impact on the dynamics of the system, whereas the spatial discretization can dramatically change them. In particular, the asymptotic Lyapunov spectrum for time refinement is achieved for CFL numbers as large as $\mathcal{O}(10^1-10^2)$, whereas the system continues to become more and more chaotic even for meshes that are much finer than the best practice for this type of flows.%Finally, we investigate the nonlinear dynamics of the flow to better understand the mechanisms that make separated flows chaotic.
%--including a quasi-DNS discretization--
%the statistics of the short-time Lyapunov exponents
%periodic orbit bifurcating into a strange attractor
%Quantitative and qualitative changes in the dynamics are investigated. 
%The existence of homoclinic tangencies in the discrete system are investigated. Finally, quasi-homoclinic tangencies are observed on or near the strange attractor of the discrete systems.
%This is important for design optimization, flow control, and uncertainty quantification as the cost and performance of chaotic sensitivity analysis methods dramatically depend on the 

%\begin{description}
%\end{description}
\end{abstract}

\pacs{Valid PACS appear here}% PACS, the Physics and Astronomy
%Use showkeys class option if keyword
\maketitle
\tableofcontents

\section{\label{s:introduction}Introduction}

%With the increase in computing power, scale-resolving turbulent simulations --such as large-eddy simulation (LES) and direct numerical simulation (DNS)-- emerge as promising approaches to improve both knowledge of complex flow physics and reliability of flow predictions. While these techniques have become essential tools in science, no comparable impact has been observed in engineering.
%The chaotic dynamics that these simulations inherit from the underlying turbulent flow are largely responsible for this lag.

Lyapunov analysis is a powerful tool to characterize dynamical systems, and the first attempts to apply it to chaotic fluid flows date back from the `90s \cite{Sirovich:1991, Keefe:1992, Pulliam:1993}. With the increase in computing power, Lyapunov analysis is gaining attention in the flow physics community \cite{Blonigan:2016:CTR,Xu:2016} as a promising approach for flow instability, vortex dynamics, and turbulence research. While the interest in flow physics lies in the Lyapunov spectrum of the actual flow, numerical algorithms compute the Lyapunov exponents of the finite-dimensional representation obtained after numerical discretization. It is therefore necessary to understand the impact of the spatial and temporal discretization on the resulting dynamics --e.g. is the spectrum of the discrete system that of the actual flow?

The Lyapunov spectrum of chaotic flow simulations also plays a key role in engineering. In particular, conventional sensitivity analysis methods break down for chaotic systems \cite{Lea:2000}, and this compromises critical tasks such as flow control, design optimization, error estimation, data assimilation, and uncertainty quantification. While a number of sensitivity analysis methods have been proposed for chaotic systems \cite{Lea:2000,Thuburn:2005,Wang:2014,Ni:2016:arXiv}, they all come at a high computational cost. This is ultimately related to the positive portion of the Lyapunov spectrum, and the cost of each method is sensitive to different aspects of it --e.g. the cost of Non-Intrusive LSS \cite{Ni:2016:arXiv} depends on the number of positive Lyapunov exponents, whereas the Ensemble Adjoint method \cite{Lea:2000} is postulated to be sensitive to the ratio of largest to smallest positive exponents \cite{Chandramoorthy:2017}--. Hence, understanding the dynamics of chaotic flow simulations, and their dependence on numerical discretization, is necessary to estimate the cost and feasibility of chaotic sensitivity analysis methods.%the various chaotic sensitivity analysis methods, drive strategic decisions about those with the most promise, and provide insight to devise superior methods.%inverse problems.%jets, wakes, separation, laminar-to-turbulent transition, shock-wave boundary-layer interactions, heat transfer, and aeromechanics instabilities.

In this paper, we investigate the Lyapunov spectrum of the separated flow around the NACA 0012 airfoil at Reynolds number $Re_{\infty} = 2,400$, Mach number $M_{\infty} = 0.2$, and angle of attack $\alpha = 20 \ \textnormal{deg.}$ Because the simulation is two-dimensional, the flow physics are different to those of three-dimensional flows. However, the moderate computational cost of this problem enable us to evaluate the impact of numerical discretization on the Lyapunov spectrum through a more comprehensive study than otherwise possible. In particular, the impact of temporal resolution ($t$-refinement), spatial resolution ($h$-refinement), and order of accuracy ($p$-refinement) are investigated.
%This motivates the choice of this case for our purpose here.
%{\color{red}{The CLVs and LEs of the two-dimensional Navier-Stokes equations on $\Omega \in \mathbb{R}^{2}$ are also CLVs and LEs of the corresponding three-dimensional Navier-Stokes equations in $\Omega \times \mathbb{R} \in \mathbb{R}^{3}$. We shall refer to the later as the quasi-2D Navier-Stokes system. This holds for a set with zero-measure in 3D, so not sure if those are indeed CLVs and LEs.}}

The paper is structured as follows. In Section \ref{s:lyapunov}, we present an overview of Lyapunov analysis. Section \ref{s:methodology} describes the methodology to discretize the Navier-Stokes equations and perform Lyapunov analysis. Numerical results are then discussed in Section \ref{s:results}. Finally, we present some concluding remarks and future work in Section \ref{s:conclusions}.

\section{\label{s:lyapunov}Lyapunov analysis}

The spatial discretization of the compressible Navier-Stokes equations yields a finite-dimensional, continuous-time, first-order dynamical system of the form
\begin{equation}
\frac{d\bm{u}_h}{dt} = \bm{f}_h(\bm{u}_h) , 
\label{e:dyn_sys}
\end{equation}
%\qquad \bm{u}_h(t_0) = \bm{u}_{h,0} , \qquad t \geq t_0, 
where $\bm{u}_h = \bm{u}_h(t)$ is an $n$-dimensional vector of state variables. In particular, $\bm{u}_h$ contains the conserved quantities (mass, momentum, total energy) at every grid point. Different meshes $h$ and numerical schemes lead to different dimensions $n$ and different dynamics $\bm{f}_h$.

For a system of the form \eqref{e:dyn_sys}, almost surely there exist scalars $\Lambda_h^1 , \Lambda_h^2 , ... , \Lambda_h^n \in \mathbb{R}$ such that, if $\Lambda_h^1 \neq \Lambda_h^2 \neq ... \neq \Lambda_h^n$, there exist vectors $\bm{\psi}_h^1(\bm{u}_h),\bm{\psi}_h^2(\bm{u}_h),...,\bm{\psi}_h^n(\bm{u}_h) \in \mathbb{R}^n$ satisfying the evolution equation \cite{Oseledets:68}
\begin{equation}
\begin{split}
\frac{d}{dt} \bm{\psi}_h^j \big( \bm{u}_h(t) \big) = & \frac{\partial \bm{f}_h}{\partial \bm{u}_h}\bigg|_{\bm{u}_h(t)} \bm{\psi}_h^j \big( \bm{u}_h(t) \big) \\ & - \Lambda_h^j \ \bm{\psi}_h^j \big( \bm{u}_h(t) \big) , \quad j=1,...,n . 
\label{e:lya_cov_def}
\end{split}
\end{equation}
$\bm{\psi}_h^j(\bm{u}_h)$ and $\Lambda_h^j$ are the so-called covariant Lyapunov vectors (CLVs) and Lyapunov exponents (LEs), respectively. We note that the CLVs depend on the state $\bm{u}_h$, whereas the LEs are a property of the system independent of $\bm{u}_h$. Also, we shall asume that the Lyapunov exponents $\Lambda_h^1 , ... , \Lambda_h^n$ are ordered from largest to smallest.%, and the CLVs have unit norm in all points of phase space.

The intuitive interpretation of Lyapunov vectors and exponents is as follows: ``Any infinitesimal perturbation $\delta \bm{u}_{h,0}$ in the direction $\bm{\psi}_h^j \big( \bm{u}(t_0) \big)$ at $t = t_0$ will remain in $\bm{\psi}_h^j \big( \bm{u}_h(t) \big)$ at all times $t \geq t_0$. Also, the magnitude of the perturbation increases or decreases at an average rate $\delta u_h (t) = \delta u_{h,0} \ \exp{ \big( \Lambda_h^j (t - t_0) \big) }$''. Hence, the magnitude and sign of the Lyapunov exponents characterize how infinitesimal perturbations to the system evolve over time. In particular, a system with $n^{+} \geq 1$ positive exponents, $\Lambda_h^{+} = \{ \Lambda_h^1 , ... , \Lambda_h^{n_+} \}$, displays chaotic dynamics. The positive exponent(s) are responsible for the ``butterfly effect'', a colloquial term to refer to the large sensitivity of chaotic systems to initial conditions. This is the case, for example, for turbulent flows as well as for many separated flows.
%Depending on the sign of the Lyapunov exponents, the dynamical system \eqref{e:dyn_sys} can be classified into four types:
%\begin{enumerate}
%\item If all the exponents are negative, the system is {\it stable} and will converge to a {\it fixed} (or {\it stationary}) {\it point} as $t \rightarrow \infty$. This is the case for steady laminar flows.
%
%\item If one LE is equal to zero and all the other exponents are negative, the system is {\it periodic} and will remain in a one-dimensional {\it limit cycle} (or {\it periodic orbit}) as $t \rightarrow \infty$. Also, $\bm{\psi}_h^1(\bm{u}_h) = \bm{f}_h(\bm{u}_h)$ is the CLV corresponding to $\Lambda_h^1 = 0$. This is the case for periodic vortex shedding.
%
%\item If $k \geq 2$ LEs vanish and all other exponents are negative, the system is {\it aperiodic} and its attractor is a $k$-dimensional torus. This is the case for aperiodic vortex shedding.
%
%\item A system with $n_{+} \geq 1$ positive exponents, $\Lambda_h^{+} = \{ \Lambda_h^1 , ... , \Lambda_h^{n_+} \}$, displays chaotic dynamics and will remain in a {\it strange attractor} as $t \rightarrow \infty$. The positive exponent(s) are responsible for the ``butterfly effect'', a colloquial term to refer to the large sensitivity that chaotic systems exhibit to initial conditions. This is the case for turbulent flows.
%
%\end{enumerate}
%a  is similar to a limit cycle, but it has at least one positive Lyapunov exponent \citep{Ginelli:2007:Lya}.

The numerical simulation of unsteady flows requires further discretizing Eq. \eqref{e:dyn_sys} in time. This yields a discrete-time first-order map
\begin{equation}
\bm{u}_{h}^{(i+1)} = \bm{f}_{h,\Delta t}(\bm{u}_{h}^{(i)}) , 
\label{e:discrete_dyn_sys}
\end{equation}
% , \qquad \bm{u}_{h}^{(i_0)} = \bm{u}_{h,0} , \qquad i \geq i_o
where $\bm{u}_{h}^{(i)}$ denotes the solution at the end of the time step $i$. The particular form of $\bm{f}_{h,\Delta t}$ depends on $\bm{f}_{h}$, that is, on the spatial discretization, as well as on the time-integration scheme and the time-step size $\Delta t$. The discrete-time Lyapunov vectors $\bm{\psi}_{h,\Delta t}^j$ and exponents $\Lambda_{h, \Delta t}^j$ of $\bm{f}_{h,\Delta t}$ are defined in an analogous way to their continuous counterparts.

\section{\label{s:methodology}Methodology}
\subsection{\label{s:numDiscr}Numerical discretization}

High-order Hybridizable Discontinuous Galerkin (HDG) and diagonally implicit Runge-Kutta (DIRK) methods are used for the spatial and temporal discretization of the compressible Navier-Stokes equations, respectively \cite{Fernandez:16a}. The HDG method, as a discontinuous Galerkin method, allows for a systematic study of the effect of the accuracy order on the Lyapunov spectrum via $p$-refinement.

\subsection{\label{s:LyaAlgh}LE algorithm}
A non-intrusive version of the algorithm by Benettin {\it et al.} \cite{Benettin:1980} is used to compute the $p \leq n$ leading Lyapunov exponents.

{\bf Original algorithm.} The original procedure in \cite{Benettin:1980} is summarized in Algorithm \ref{originalAlg}. If the time integrals in Steps No. 5 and 6 of the algorithm are computed exactly, an estimator of the $p$ leading continuous-time LEs $\hat{\Lambda}_h$ of $\bm{f}_h$ are obtained. If the time integrals are approximated using a numerical method, as it is the case in practice, the algorithm computes an estimator of the $p$ leading discrete-time Lyapunov exponents $\hat{\Lambda}_{h,\Delta t}$ of $\bm{f}_{h,\Delta t}$.
%This is a stable algorithm to {\it almost surely} compute unbiased estimators of the $p$ leading LEs and CLVs.

\vspace{3mm}

\begin{algorithm}[H]
\label{originalAlg}
 \KwData{Initial condition $\bm{u}_{h,0}$, number of exponents to compute $p$, length of each time segment $T_s$, and number of time segments $K$.}
 \KwResult{Estimators of the $p$ largest LEs $\hat{\Lambda}_h^j , \ j=1,...,p$.}
 1. Set $t_0 = 0$ and $\bm{u}_{h}(t_0) = \bm{u}_{h,0}$. \\
 2. Compute an $n \times p$ random matrix
 $$V^{\langle 0 \rangle} \sim \big [ \mathcal{U} (0, 1) \big ] ^{n \times p} . $$ \\
 3. Compute the reduced QR decomposition
 $$Q^{\langle 0 \rangle} R^{\langle 0 \rangle} = V^{\langle 0 \rangle} . $$
 \For{$i = 1$ \KwTo $K$}{
    \indent 4. Set $t_{i} = t_{i-1} + T_s$. \\% where $T_s^*$ is the length of the time segments $[t_i^*,t_{i+1}^*]$.
    \indent 5. Time integrate the dynamical system \eqref{e:dyn_sys} from $t_{i-1}$ to $t_i$ using the initial condition $\bm{u}_{h}^{\langle i-1 \rangle} = \bm{u}_{h}(t_{i-1})$. \\
    %$$\frac{d \bm{u}_{h}}{d t} = \bm{f}_h (\bm{u}_{h}) , \qquad \bm{u}_{h}(t_{i-1}) = \bm{u}_{h}^{(i-1)} , \qquad t \in [t_{i-1}, t_i] . $$ \\
   	\indent 6. Time integrate the tangent equation \eqref{e:tangentEq} from $t_{i-1}$ to $t_i$ for each of the $p$ initial conditions given by the columns of $Q^{\langle i-1 \rangle}$ using the reference trajectory $\bm{u}_{h}$ computed in Step No. 5,
%\footnote{$T_h$ denotes the tangent map from $t_1$ to $t_2$, i.e. $\bm{v}_h (t_2) = T_{h,\Delta t} \big( \bm{v}_h (t_1),t_1,t_2 \big)$.}
\begin{equation}
\frac{d \bm{v}_j}{dt} = \frac{\partial \bm{f}_h}{\partial \bm{u}_h}\bigg|_{\bm{u}_h(t)} \bm{v}_j , \qquad \bm{v}_j(t_{i-1}) = \bm{q}_j^{\langle i-1 \rangle} , 
\label{e:tangentEq}
\end{equation}
   	%$$V^{\langle i \rangle} = T_h(Q^{\langle i-1 \rangle} ; \bm{u}_{h} , t_{i-1}, t_{i}) . $$
for $j=1,...,p$, and set $\bm{v}^{\langle i \rangle}_j = \bm{v}_j(t_{i})$. Here, $\bm{q}^{\langle i \rangle}_j$ and $\bm{v}^{\langle i \rangle}_j$ denote the $j$-th column of $Q^{\langle i \rangle}$ and $V^{\langle i \rangle}$. \\
   	\indent 7. Compute the reduced QR decomposition
   	$$ Q^{\langle i \rangle} R^{\langle i \rangle} = V^{\langle i \rangle} . $$ \\
   }
   8. Compute
\begin{equation}
\label{lyaEstimators}
\hat{\Lambda}_h^j = \frac{1}{t_K-t_0} \sum_{i=1}^{K} \log{|R^{\langle i \rangle}_{jj}|} . 
\end{equation}
 \caption{Original algorithm by Benettin {\it et al.} \cite{Benettin:1980} to compute LEs. 
 The superscript $\langle i \rangle$ denotes the solution at the end of the time segment $i$.}
\end{algorithm}

\vspace{3mm}

%We shall refer to the intervals $[t_i^*,t_{i+1}^*]$ as time segments.
%The intuition behind the algorithm is as follows. First, an invertible linear mapping, such as $T_h(\bm{q}_{i};t_i^*,t_{i+1}^*)$, maps a $p$-dimensional vector subspace $V_{i}$ onto another $p$-dimensional vector subspace $V_{i+1}$. More importantly, the rate of growth of the $j$-dimensional Lebesgue measure $\mu_j$ of an open set $A \subseteq V$ of dimension $j$, that is, $\mu_j \big( T_h(A) \big) / \mu_j (A)$, is almost surely independent of the set $A$. Hence, by periodically orthonormalizing $V(t)$, the algorithm above leads to a numerically stable procedure to compute the rates of growth of the subspaces spanned by the $j = 1 , ... , p$ first columns of $V_0$. For almost any\footnote{Except for the zero-measure collection of sets that are orthogonal to one or more of the $p$ leading covariant Lyapunov vectors $\bm{\psi}^j \big( \bm{u}_h(t_0^*) \big) , \ j = 1 , ... p$.} $V_0$, this corresponds to the product of the $j$ leading Lyapunov exponents.

{\bf Modified algorithm.} Since the original algorithm requires the integration of the tangent equation \eqref{e:tangentEq} in Step No. 6, it cannot be used with existing computational fluid dynamics (CFD) solvers without modification of the source code. In the spirit of making the algorithm non-intrusive, we approximate the tangent map \eqref{e:tangentEq} by finite differences. In particular, let $\bm{u}_{h}^{\langle i \rangle} = \bm{g}_{h,\Delta t}(\bm{u}_{h}^{\langle i-1 \rangle} ; \ T_s)$ denote the Navier-Stokes map over the time segment $[t_{i-1},t_i]$ of length $T_s$ computed by a CFD code starting from the initial condition $\bm{u}_{h}^{\langle i-1 \rangle}$ at $t_{i-1}$. We then replace Step No. 6 in Algorithm \ref{originalAlg} by
\begin{equation}
\begin{split}
\bm{v}^{\langle i \rangle}_j \approx & \frac{1}{\epsilon} \Big[ \bm{g}_{h,\Delta t} \big( \bm{u}_{h} (t_{i-1}) + \epsilon \ \bm{q}^{\langle i-1 \rangle}_j ; \ T_s \big) \\ &- \bm{g}_{h,\Delta t} \big( \bm{u}_{h} (t_{i-1}) ; \ T_s \big) \Big] , \qquad j = 1, ..., p , 
\end{split}
\end{equation}
Here, $T_s$ is small enough such that $|\bm{g}_{h,\Delta t} \big( \bm{u}_{h} (t_{i-1}) + \epsilon \ \bm{q}^{\langle i-1 \rangle}_j ; \ T_s \big) - \bm{g}_{h,\Delta t} \big( \bm{u}_{h} (t_{i-1}) ; \ T_s \big)| \ll 1$, and $\epsilon$ satisfies $\epsilon_{tol} \ll \epsilon \ll 1$, where $\epsilon_{tol}$ denotes the tolerance of the solver for the nonlinear system of equations arising from the DIRK discretization.

\section{\label{s:results}Numerical results}

%\subsection{Motivation}
%
%IN THIS SENSE (MOTIVATION), IS IT TRUE / CAN WE SHOW THAT CLVs AND LEs OF 2D FLOW ARE ALSO CLVs AND LEs OF 3D FLOW?
%
%The original study by Pulliam \cite{TBD} revealed that different algorithms to compute Lyapunov exponents may predict very different exponents. In particular, Algorithm XX predicted $\lambda_1 = 0.97$ ($\lambda_1 = 2.63$), whereas Algorithm XX yielded $\lambda_1 = 5.4$ ($\lambda_1 = 6.0$) at M=0.2 and Re=2k (3k), although the data from the same numerical simulation was used in both algorithms.
%
%A preliminary study on the leading Lyapunov exponent with different flow solvers confirmed the large impact of numerics on the Lyapunov spectrum\footnote{We note that, unlike elsewhere in this paper, the Lyapunov exponents above are nondimensionalized with respect to the freestream velocity $u_{\infty}$ in order to eliminate the effect of the Mach number and allow comparison between the Lyapunov exponents at different Mach numbers.}:
%\begin{itemize}
%\item Reynolds (Re=2.4k, M=0.2): 0.213
%\item Eddy (Re=2.4k, M=0.2): 0.35
%\item FUN3D (Re=10k, M=0.1): 0.173
%\end{itemize}
%All these exponents are much smaller than those computed by Pulliam \cite{TBD}. This confirms that the relationship between numerics and chaotic dynamics is complicated. In particular, factors such as the numerical scheme, order of accuracy, mesh resolution, and time-step size seem to have a major influence on the underlying discrete dynamics.  Next, we examine each of these factors independently.
%%very large impact of the discretization on the Lyapunov exponents.

\subsection{Case description}

We consider the two-dimensional, separated flow around the NACA 0012 airfoil at Reynolds number $Re_{\infty} = u_{\infty} c /\nu = 2400$, Mach number $M_{\infty} = u_{\infty} / a_{\infty} = 0.2$, and angle of attack $\alpha = 20 \ \textnormal{deg}$. Here, $u_{\infty}$, $a_{\infty}$, $\nu$ and $c$ denote the freestream velocity, freestream speed of sound, kinematic viscosity, and airfoil chord, respectively. The computational domain is partitioned using isoparametric triangular elements, and the outer boundary is located 10 chords away from the airfoil. A non-slip, adiabatic wall boundary condition is imposed on the airfoil surface, and a characteristics-based, non-reflecting boundary condition is used on the outer boundary.
%The impact of spatial resolution ($h$-refinement), accuracy order ($p$-refinement), and temporal resolution ($t$-refinement) on the Lyapunov spectrum is investigated. %A snapshot of the Mach number field for this flow is shown in Fig. \ref{TBD}.
%The moderate computational cost of this problem allows us to perform an exhaustive analysis that would not be possible otherwise. This, together with the extensive analysis of this flow in \cite{Pulliam:1993}, motivates the choice of this case for the purpose here.
%Despite the low Reynolds number, the flow shows chaotic dynamics due to the massive separation induced by the large angle of attack.
%While two-dimensional turbulence is simpler and fundamentally different from three-dimensional turbulence, this case has been chosen due to the extensive analysis of the chaotic behavior of this flow in \cite{Pulliam:1993}. 
%Also, it is important from a theoretical perspective, and 2D results can be thougth of as best-case scenario for 3D trends.

\subsection{\label{s:tRef}Effect of time resolution: $t$-refinement}
We analyze the effect of the time-step size on the Lyapunov spectrum of the discrete system $\bm{f}_{h,\Delta t}$. In particular, the continuous-time system $\bm{f}_h$ associated to a fourth-order discretization (i.e. $p=3$) with 115,200 degrees of freedom (DOFs) is time-integrated using the time steps $\Delta t = (0.0400$, $0.0200$, $0.0100$, $0.0050$, $0.0025) \ c / a_{\infty}$. These correspond to maximum CFL numbers of 59.94, 29.97, 14.98, 7.49, and 3.75. We emphasize that the time-step size affects the discrete-time map $\bm{f}_{h,\Delta t}$ but does not change $\bm{f}_h$.%while $h$- and $p$-refinement leaded to different continuous-time dynamical systems $\bm{f}_h$, 
%The initial condition $\bm{u}_{h,0}$ in the LE and CLV algorithms is the same for all time-step sizes.
% Minimum element height: 6.673889873575041e-04

\begin{figure}[H]
\centering
\includegraphics[width=0.5\textwidth]{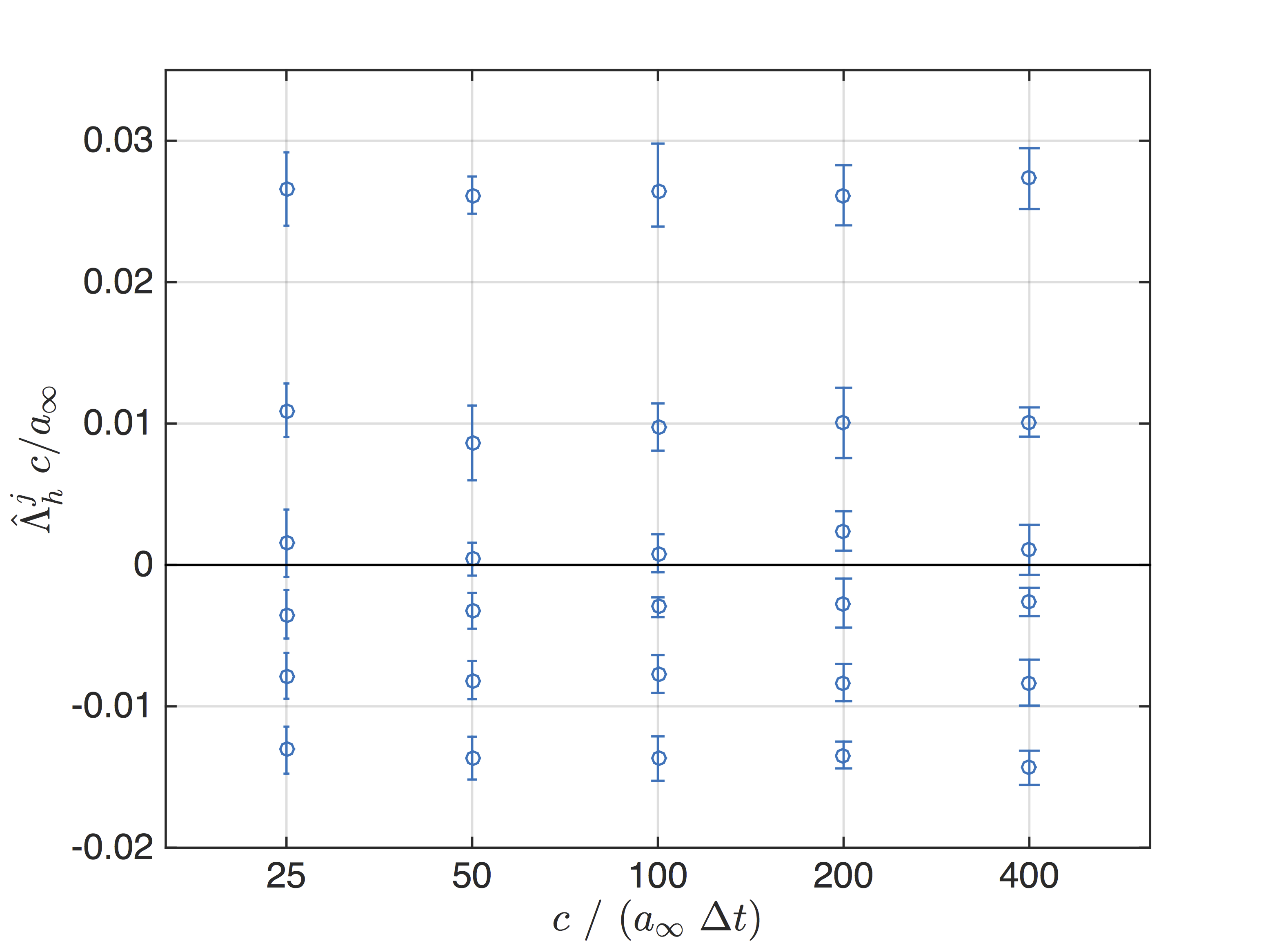}
\caption{\label{f:lyaExp_tRef} 90\% confidence intervals of the six leading Lyapunov exponents in the $t$-refinement study.}
\end{figure}

Figure \ref{f:lyaExp_tRef} shows 90\% confidence intervals of the six leading Lyapunov exponents for the time-step sizes considered.  The confidence intervals are computed from the sample variance of $\log{|R^{\langle i \rangle}_{jj}|}$ in Eq. \eqref{lyaEstimators} and the Central Limit Theorem. From this figure, the time-step size in the range considered does not have a significant impact on the leading exponents of $\bm{f}_{h,\Delta t}$. First, this gives us confidence that the time steps considered suffice for the discrete-time Lyapunov exponents to approximate those of the continuous-time system, i.e. $\Lambda_{h,\Delta t}^j \approx \Lambda_h^j$. For this reason, we shall refer to $\bm{f}_h$ and $\Lambda_h$, instead of $\bm{f}_{h,\Delta t}$ and $\Lambda_{h,\Delta t}$, in the remainder of the paper. Second, the asymptotic spectrum of the discrete-time map as $\Delta t \rightarrow 0$ is achieved with CFL numbers $\mathcal{O}(10-100)$ that are larger than those used in engineering practice. This is attributed to these time-step sizes being sufficiently small to resolve the vortical structures that are responsible for the chaotic dynamics. However, if $\Delta t \gg h_{sep} / u_{\infty}$, where $h_{sep}$ denotes the element size in the separated region, the discrete-time map might not accurately reproduce the continuous-time system, and therefore $\Lambda_{h,\Delta t}^j \not\approx \Lambda_h^j$. This has indeed been observed in \cite{Ozkaya:2016} for the numerical integration of stiff ODEs with inadequate time steps.
%indicates that the time-step size has a minor impact on the Lyapunov spectrum provided it is sufficiently small to resolve the main dynamics of the continuous-time system. 

\subsection{\label{s:hRef}Effect of spatial resolution: $h$-refinement}
\begin{figure*}
\centering
\includegraphics[width=0.7\textwidth]{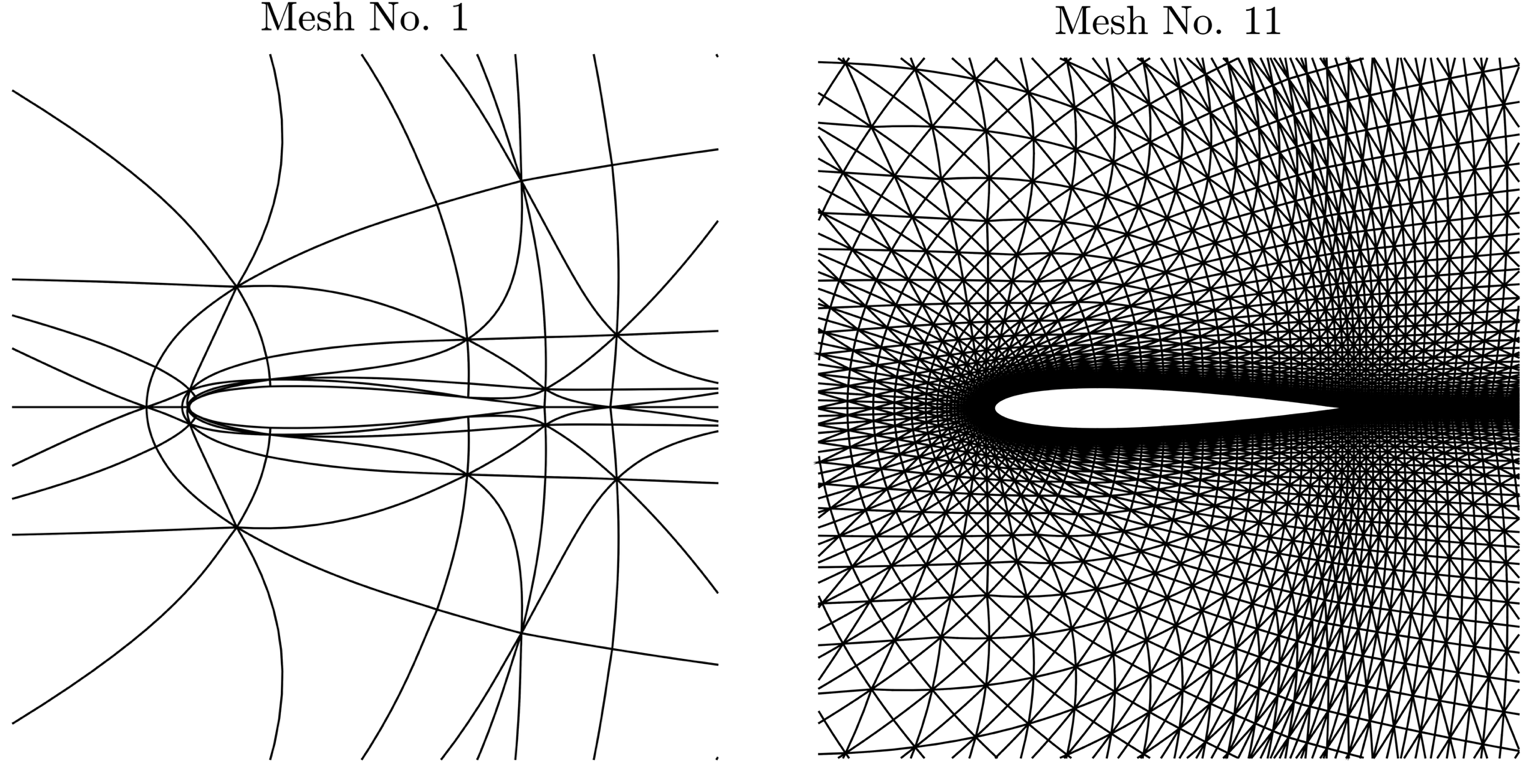}
\caption{\label{meshes} Coarsest and finest high-order meshes used in the $h$-refinement study.}
\end{figure*}

%All the results in this paper are expressed in terms of nondimensional time $t^*$. The freestream speed of sound $a_{\infty}$ and the airfoil chord $c$ are used as velocity and length scales, respectively, so that $t^* = t a_{\infty} / c$.
In this section, we examine the effect of the spatial resolution on the number and magnitude of positive exponents. To that end, the Lyapunov spectrum is computed for eleven O-meshes, each of them $2^{1/3}$ times finer per direction than the previous one. The number of DOFs uniformly increases in logarithmic scale from 7,200 (mesh No. 1) to 726,240 (mesh No. 11). Meshes No. 1 and 11 are shown in Fig. \ref{meshes}. We note that mesh No. 1 is intended to be pathologically coarse to analyze how the system behaves for very under-resolved meshes.%in the under-resolution limit.

%Example of three computational meshes used in the $h$-refinement study.

The discretization scheme and time-step size are kept constant to analyze the effect of spatial resolution only. In particular, fourth-order ($p=3$) HDG and third-order DIRK methods are used for the spatial and temporal discretization, respectively, and the time-size is set to $\Delta t = 0.05 \ c / a_{\infty}$. A run up time of $2,000 \ c / a_{\infty}$ is used to drive the system to the attractor, and the LE algorithm is then applied for $K=12,000$ time segments each of length $T_s = c / a_{\infty}$. Figure \ref{f:leadingLE_NACA0012} shows the 14 leading LEs for the discretizations considered, whereas Table \ref{t:dimAttractor} collects the Kaplan-Yorke dimension $D_{KY}$ \cite{Eckmann:1985} of the $h$-family of attractors. From these results, several remarks follow:
%and Fig. \ref{CIofLeadingLE_NACA0012} displays the 90\% confidence intervals for the four leading exponents.
%This time step is small compared to the element size so that spatial discretization error dominates.
%As discussed in Section \ref{s:pRef}, the latter may not be a good approximation for the near-zero exponents, and the corresponding CIs may suffer from significant errors.
\begin{itemize}
%\item The magnitude of the leading LE and the number of positive exponents are not strictly increasing functions of the spatial resolution. In particular, both reach a minimum for the resolution corresponding to mesh No. 5. We note that discretization No. 5 has no positive LEs and is indeed periodic, and that no discretization results in stable dynamics. In other words, for the $h$-family of discrete dynamical systems considered here, a periodic orbit bifurcates into a strange attractor above and below $\sim h_5$. A time refinement study confirmed that the continuous-time system associated to discretization No. 5 --and not only the discrete-time map-- is indeed periodic.
\item The magnitude of the leading LE and the number of positive exponents increase above a spatial resolution threshold $h^*$, corresponding to mesh No. 5. That is, the discrete system becomes more chaotic above this resolution as the mesh is refined. This is attributed to the fact that more vortical structures, which are responsible for the chaotic dynamics of the flow, are resolved as the numerical resolution is increased.%We postulate this is true for any chaotic system discretized using a convergent scheme.%; although the resolution threshold is scheme-dependent.
%In this study, this threshold corresponds to mesh No. 5.

\begin{figure}[H]
\centering
\includegraphics[width=0.5\textwidth]{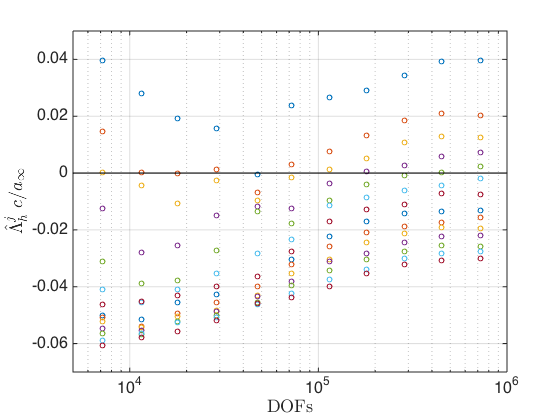}
\caption{\label{f:leadingLE_NACA0012} 14 leading Lyapunov exponents of the $h$-family of dynamical systems.}
\end{figure}

\item Below the resolution threshold $h^*$, the discrete system poorly reproduces the dynamics of the continuous system, and this results in spurious dynamics. Here, spurious periodicity and chaoticity are observed. (No discretization results in stable dynamics.) Discretization No. 5, for example, has no positive LEs and is periodic. A time refinement study confirmed that the continuous-time system associated to this discretization $\bm{f}_{h_5}$ --and not only the discrete-time map $\bm{f}_{h_5,\Delta t}$-- is indeed periodic. Hence, for the $h$-family of discrete dynamical systems considered here, a periodic orbit bifurcates into strange attractors above and below $\sim h_5$.

We hypothesize this is a numerical artifact and therefore discretization dependent. For example, spurious chaoticity may not be observed in methods with high numerical dissipation, such as first-order schemes. For these methods, stable dynamics with a fixed point could be obtained instead with a very coarse mesh. %with a fixed point {\it close} to the average state of the continuous system 
\item An approximately zero exponent $\Lambda_h \approx 0$ is present in all discretizations. Theoretical results show that $\Lambda_h = 0$ with $\bm{\psi}_h = \bm{f}_h(\bm{u}_h)$ for periodic and chaotic systems. This is expected to be such a exponent, and the error is attributed to the variance of the estimator and, to a lessen extent, the approximation $\Lambda_{h,\Delta t} \approx \Lambda_h$.
%This is consistent with theoretical results for periodic and chaotic systems, and corresponds to perturbations in the $\bm{f}_h(\bm{u}_h)$ direction in phase space.
\item The positive Lyapunov exponents are created from bifurcations of the $\Lambda_h = 0$ exponent at discrete mesh resolutions.
\end{itemize}
%As expected, the dimension of the attractor increases above and below resolution No. 5. %We recall that the Kaplan-Yorke dimension is smaller than the capacity, fractal or Hausdorff dimension $D_{KY} < D_{cap}$. -> Not sure about this.

%\begin{figure}[H]
% \centering
% \subcaptionbox{ \label{leadingLE_NACA0012}}
% [7.4cm]{\includegraphics[scale = .40]{leadingLE_NACA0012.png}}
% \hfill \subcaptionbox{ \label{CIofLeadingLE_NACA0012}}
%  [7.4cm]{\includegraphics[scale = .40]{CIofLeadingLE_NACA0012.png}}
%  \caption{(a) 14 leading Lyapunov exponents of the $h$-family of dynamical systems, and (b) 90\% confidence intervals for the 4 leading exponents.}\label{LE_NACA0012}
% \end{figure}
%\begin{table}[htbp]
%\centering
%\begin{tabular}{cccc}
%%\toprule
%Mesh No. & $D_{KY}$ & Mesh No. & $D_{KY}$ \\
%%\midrule
%\hline
%1 & 4.7306 & 7 & 5.3711 \\
%2 & 2.1409 & 8 & 7.7215 \\
%3 & 2.6830 & 9 & 8.2739 \\
%4 & 2.0117 & 10 & 9.1949 \\
%5 & 1.0000 & 11 & 10.9038 \\
%6 & 3.2899 &  & \\
%\hline
%%\bottomrule
%\end{tabular}
%\caption{\label{t:dimAttractor} Kaplan-Yorke dimension \cite{Eckmann:1985}, $D_{KY}$, of the $h$-family of attractors.}
%\end{table}
\begingroup
%\squeezetable
\begin{table*}
\centering
%\begin{ruledtabular}
\begin{tabular}{cccccccccccc}
\hline
Discretization No. & 1 & 2 & 3 & 4 & 5 & 6 & 7 & 8 & 9 & 10 & 11  \\
\hline
$St$ & 0.24 & 0.23 & 0.26 & 0.24 & 0.26 & 0.26 & 0.26 & 0.26 & 0.26 & 0.26 &  0.26 \\
\hline
$D_{KY}$ & 4.73 & 2.14 & 2.68 & 2.01 & 1.00 & 3.29 & 5.37 & 7.72 & 8.27 & 9.19 & 10.90  \\
\hline
\end{tabular}
%\end{ruledtabular}
\caption{\label{t:dimAttractor} Strouhal number $St$ and Kaplan-Yorke attractor dimension $D_{KY}$ \cite{Eckmann:1985} of the $h$-family of discretizations.}
\end{table*}
\endgroup
%\begin{figure}[h!]
%\centering
%\includegraphics[width=0.5\textwidth]{attractorDimension.png}
%\caption{\label{f:dimAttractor} Kaplan-Yorke dimension \cite{Eckmann:1985}, $D_{KY}$, of the $h$-family of attractors.}
%\end{figure}
%\begin{figure}[h!]
%\centering
%\includegraphics[width=0.5\textwidth]{leadingLE_NACA0012.png}
%\caption{\label{leadingLE_NACA0012} Leading Lyapunov exponents.}
%\end{figure}
%
%\begin{figure}[h!]
%\centering
%\includegraphics[width=0.8\textwidth]{CIofLeadingLE_NACA0012.png}
%\caption{\label{CIofLeadingLE_NACA0012} 90\% confidence interval of four leading Lyapunov exponents.}
%\end{figure}
%Figure \ref{convMeshNo5} shows the convergence history of LEs and 90\% CIs for Mesh No. 5. 

\begin{figure}[H]
 \centering
 \subcaptionbox{Mesh No. 5 \label{f:cl_cd_trace_meshNo5}}
 {\includegraphics[width=0.45\textwidth]{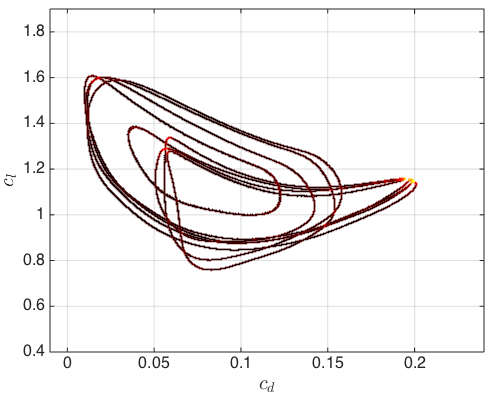}}
 \hfill \subcaptionbox{Mesh No. 7 \label{f:cl_cd_trace_meshNo7}}
 {\includegraphics[width=0.45\textwidth]{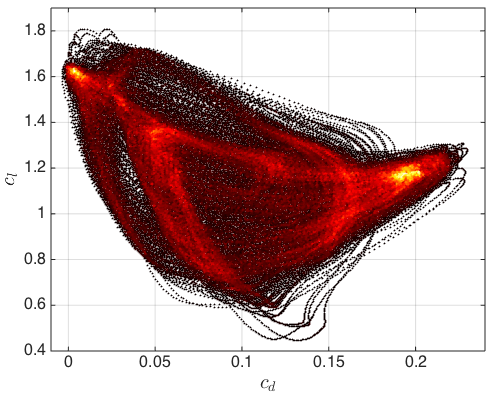}}
  \caption{$(c_d,c_l)$ trace over a time interval $\Delta t = 20,000 \ c / a_{\infty}$. The dots are colored by probability density function in $(c_d,c_l)$ space.}\label{f:cl_cd_trace}
 \end{figure}

%The Fourier transform of the lift coefficient $c_l$, drag coefficient $c_d$, and static pressure at $(x,y) = (0.9,0.2)$ for discretizations No. 5 and 7 are displayed in Fig. \ref{DFT_meshNo5_7}.
The trace of drag $c_d$ and lift $c_l$ coefficients over a time interval $\Delta t = 20,000 \ c / a_{\infty}$ is shown in Fig. \ref{f:cl_cd_trace}, where the dots are colored by probability density function (PDF) in $(c_d,c_l)$ space. Despite the chaotic dynamics of discretization No. 7, we note that the PDF resembles the periodic trace of mesh No. 5. The period for this discretization, $T_5 \approx 75.70 \ c / a_{\infty}$, is four times the dominant vortex shedding period. The Strouhal number for the $h$-family of discretizations are collected in Table \ref{t:dimAttractor}.%{\color{red}{Look at a movie}}
%This figure nicely illustrates that discretization No. 5 yields periodic dynamics, whereas the dynamics of mesh No. 7 are chaotic.
% 75.70 (DFT) 56.8 (trace)$.
%St:
%1: 0.2435
%2: 0.2332
%3: 0.2586
%4: 0.2367
%5: 0.2650
%7: 0.2626

%\begin{figure}[h!]
%\centering
%\includegraphics[width=1.0\textwidth]{DFT_meshNo5_7.png}
%\caption{\label{DFT_meshNo5_7} Discrete Fourier Transform of lift coefficient, drag coefficient, and pressure at $(x,y) = (0.9,0.2)$ for discretizations No. 5 (top) and 7 (bottom).}
%\end{figure}

Next, we investigate if an asymptotic Lyapunov spectrum is achieved with the numerical resolutions that can be afforded in engineering practice. To this end, we consider a fourth-order discretization with 2,880,000 degrees of freedom. This is vastly more than the best-practice meshes for this type of flows. The 90\% confidence interval of the leading Lyapunov exponent for this discretization is $\Lambda_h^1 \ c / a_{\infty} = 0.04732 \pm 0.005275$. Hence, the system continues to become more chaotic even for this discretization.%when the numerical resolution is close to the DNS regime.
%This grid count is such that the equivalent resolution in three dimensions is near the limit of what can be resolved with current supercomputers.

While an asymptotic Lyapunov spectrum as $h \rightarrow 0$ was obtained for simpler partial differential equations in other studies \cite{Takeuchi:2011}, this result shows that such an asymptotic spectrum --if exists-- is difficult to achieve in practice even for simple flows. This is in contrast to the results for the time discretization in Section \ref{s:tRef}.

%We postulate, however, that $\lim_{h \rightarrow 0^+} {\Lambda^+_h}$ exists and is equal to $\Lambda^+$ for any convergent scheme.{\color{red}{ Could it be a superset of $\Lambda^+$?}} That is, the set of discrete positive exponents asymptotes to the set of continuous positive exponents as the spatial resolution increases. This asymptotic behavior may be difficult to observe in practice due to limited computational resources, as demonstrated here. What is more, the lower bound in the discretization error due to finite precision arithmetic may hinder achieving the asymptotic spectrum even with unlimited computational resources.%Numerical results showed that the discrete system of the NACA 0012 airfoil at $Re_{\infty} = 2,400$ and $\alpha = 20 \ \textnormal{deg.}$ continues to become more chaotic even when the resolution is close to the DNS regime.

%The validity and generality of the hypotheses in this section is to be examined in other scenarios, such as different flow conditions and CFD solvers.%While this is the subject of current work, we encourage other groups to examine the hypotheses above.

\subsection{\label{s:pRef}Effect of spatial accuracy order: $p$-refinement}
Finally, we investigate the effect of the accuracy order of the spatial discretization on the dynamics of $\bm{f}_h$. To this end, third-, fourth-, and fifth-order HDG schemes (i.e. ($p=\{2,3,4\}$) are considered. The DIRK(3,3) method with $\Delta t = 0.05 \ c / a_{\infty}$ is used for the time integration, and the number of degrees of freedom is 115,200 in all cases. This corresponds to resolution No. 7 in the $h$-refinement study.

\begin{figure}
\centering
\includegraphics[width=0.5\textwidth]{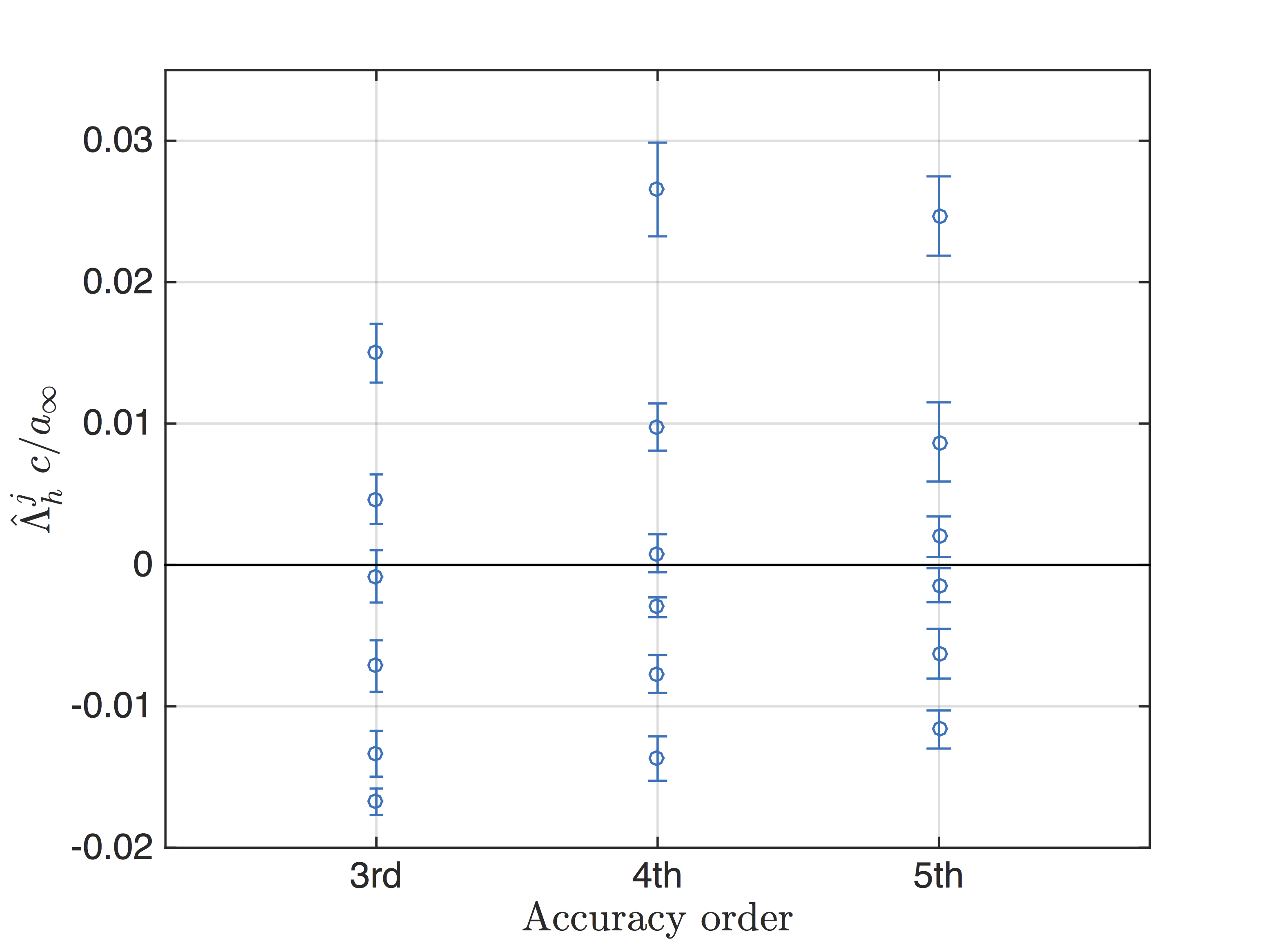}
\caption{\label{f:lyaExp_pRef} 90\% confidence intervals of the six leading Lyapunov exponents in the $p$-refinement study.}
\end{figure}

Figure \ref{f:lyaExp_pRef} shows 90\% confidence intervals of the six leading Lyapunov exponents for the accuracy orders considered. From this figure, the negative LEs get closer to zero, that is, perturbations along stable directions decay more slowly, as the accuracy order increases. This is attributed to the lower numerical dissipation of high-order methods. Also, the fourth- and fifth-order methods lead to larger positive exponents than the third-order scheme --i.e. perturbations along unstable directions get more rapidly amplified. The 90\% condiference intervals for the fourth- and fifth-order discretizations overlap and it is not possible to conclude which scheme results in more chaotic dynamics.
%The unstable modes show more complex behavior. In particular, the fourth-, fifth-, and third-order discretizations lead to the largest positive exponents --i.e. perturbations along unstable directions get more rapidly amplified--, in this order. We postulate this is due to either the variance of the estimators (the 90\% condiference intervals do overlap) or the different nature of the leading-order error term of each scheme.%{\color{red}{ The $p=4$ discretization is still running to check if the estimators shift up.}}

\section{\label{s:conclusions}Conclusions}

%We investigated the Lyapunov spectrum of separated flows and their dependence on the numerical discretization. The chaotic flow around the NACA 0012 airfoil at low Reynolds number and large angle of attack was consider to that end.
We investigated the impact of the numerical discretization on the Lyapunov spectrum of chaotic, separated flow simulations. Numerical results showed that the time discretization has a small effect on the Lyapunov spectrum for the time-step sizes typically used in CFD practice. In particular, the asymptotic spectrum as $\Delta t \rightarrow 0$ was achieved for CFL numbers $\mathcal{O}(10^1-10^2)$. The spatial discretization, however, was shown to dramatically change the dynamics of the system. First, the discretized system poorly reproduced the dynamics of the flow, and spurious dynamics were observed, below some spatial resolution threshold. Second, the discrete system continued to become more and more chaotic even with finer meshes than the best practice for this type of flows. This indicates that the asymptotic Lyapunov spectrum as $h \rightarrow 0$, if exists, is difficult to achieve in practice even for simple flows.%, and illustrates the need 
\begin{acknowledgements}
The authors acknowledge AFOSR Award 14RT0138 under Dr. Fariba Fahroo and Dr. Jeanluc Cambrier, and Stanford CTR Summer Program 2016. The first author also thanks ``la Caixa'' Foundation for the Graduate Studies Fellowship that support his work.
%We also gratefully acknowledge XX and the XX for partially supporting this effort.
%The work of the XX author is supported by XX.
\end{acknowledgements}

%\paragraph{Eliding repeated information}
%When a reference is merged, some of its fields may be elided: for example, 
%when the author matches that of the previous reference, it is omitted. 
%If both author and journal match, both are omitted.
%If the journal matches, but the author does not, the journal is replaced by \emph{ibid.},
%as exemplified by Ref.~[\onlinecite{epr}]. 
%These rules embody common editorial practice in APS and AIP journals and will only
%be in effect if the markup features of the APS and AIP Bib\TeX\ styles is employed.
%
%\paragraph{The options of the cite command itself}
%Please note that optional arguments to the \emph{key} change the reference in the bibliography, 
%not the citation in the body of the document. 
%For the latter, use the optional arguments of the \verb+\cite+ command itself:
%\verb+\cite+ \texttt{*}\allowbreak
%\texttt{[}\emph{pre-cite}\texttt{]}\allowbreak
%\texttt{[}\emph{post-cite}\texttt{]}\allowbreak
%\verb+{+\emph{key-list}\verb+}+.

\end{document}